# First-principles Study On The Electronic And Optical Properties Of Cubic ABX$_3$ Halide Perovskites

Li Lang, Ji-Hui Yang, Heng-Rui Liu, H. J. Xiang* and X. G. Gong*

Key Laboratory for Computational Physical Sciences (MOE), State Key Laboratory of Surface Physics, and Department of Physics, Fudan University, Shanghai 200433, China

## ABSTRACT

The electronic properties of ABX$_3$ (A = Cs, CH$_3$NH$_3$, NH$_2$CHNH$_2$; B = Sn, Pb; X = Cl, Br, I) type compounds in the cubic phase are systematically studied using the first-principles calculations. We find that these compounds have direct band gaps at R point where the valance band maximum is an anti-bonding state of B *s*-X *p* coupling, while the conduction band minimum is a non-bonding state with B *p* characters. The chemical trend of their properties as A or B or X varies is fully investigated, which is of great importance to understand and optimize this kind of solar cell materials. We find that: (i) as the size of A increases, the band gap of ABX$_3$ will increase; (ii) as B varies from Sn to Pb, the band gap of ABX$_3$ will increase; and (iii) as X ranges from Cl to Br to I, the band gap will decrease. We explained these trends by analyzing their band structures. Furthermore, optical properties of the ABX$_3$ compounds are investigated. Our calculations show that taking into account the spin-orbit coupling effect is crucial for predicting the accurate band gap of these halide perovskites. We predict that CH$_3$NH$_3$SnBr$_3$ is a promising material for solar cells absorber with a perfect band gap and good optical absorption.

## I. INTRODUCTION

Recently, I-IV-VII$_3$ materials (noted as ABX$_3$) with halide perovskite structures have drawn wide interest for their potential application as solar-cell absorbers,[1-5] topological insulators,[6] and even superconductors.[7] Experimentally, Chung *et al.* reported a new type of all-solid-state, inorganic solar cell system that consists of the p-type direct band gap semiconductor CsSnI$_3$ and n-type nanoporous TiO$_2$ with the dye N719, showing a conversion efficiency of up to 10.2%.[1] Lee *et al.* also reported a low-cost, solution-processable solar cell, based on a highly crystalline perovskite absorber (CH$_3$NH$_3$PbI$_2$Cl) with intense visible to near-infrared absorptivity, that has a power conversion efficiency of 10.9% in a single-junction device under simulated full sunlight.[2] More recently, Burschka *et al.* achieved solution-processed photovoltaic cells based on CH$_3$NH$_3$PbI$_3$/TiO$_2$ with unprecedented power conversion efficiencies (approximately 15%) and high stability equal to or even greater than those of today's best thin-film photovoltaic devices.[3] All of these experiments revealed that this kind of materials could be important solar cell absorbers. However, theoretical studies on this kind of material were rather incomplete.[8-11] For example, Borriello *et al.* studied the structural and electronic properties of tin based ABX$_3$ compounds,[9] Chang *et al.* also studied the structural and electronic properties of lead based ABX$_3$ compounds,[10] and Murtaza *et al.* studied the structural and optoelectronic properties of cubic peroviskites CsPbX$_3$ (X = Cl, Br, I).[11] All these studies adopted the local (or semi-local) density approximation, which severely underestimates the band gaps. Furthermore, the spin-orbit coupling effect in these compounds with heavy elements was not taken into account in these studies. Given the importance of these halide perovskites, a systematic and accurate investigation on the ABX$_3$ type compounds is desirable, which might be very important to understand and optimize this kind of solar cell absorption materials.

In this paper, we systematically investigate the electronic properties of series of ABX$_3$ (A = Cs, CH$_3$NH$_3$, NH$_2$CHNH$_2$; B = Sn, Pb; X = Cl, Br, I) type compounds using first-principles calculations, and analyze the chemical trends of their properties as A or B or X varies. We find that: (i) as the size of A increases, the band gap of

ABX$_3$ will increase; (ii) as B changes from Sn to Pb, the band gap of ABX$_3$ will increase; and (iii) as X changes from to Cl to Br to I, the band gap will decrease. We also study the optical properties of these materials and propose CH$_3$NH$_3$SnBr$_3$ might be a promising material for solar cells absorber with a perfect band gap and good optical absorption.

## II. CALCULATION METHODS

Our calculations are performed using the Vienna ab initio simulation package (VASP)[12] in the framework of density functional theory (DFT). The projector augmented-wave (PAW) pseudopotentials[13] are used with an energy cutoff of 500 eV for the plane-wave basis functions. The Monkhorst-Pack k-point mesh[14] of $8 \times 8 \times 8$ is employed and is further increased to $12 \times 12 \times 12$ in calculating the optical properties. The lattice vectors and atomic positions are optimized according to the guidance of atomic forces, with a criterion that requires the calculated force on each atom smaller than 0.01 eV/Å. For the exchange-correlation functional, the generalized gradient approximation (GGA) of Perdew-Burke-Ernzerhof (PBE)[15] is used to relax the structural parameters and to calculate the imaginary dielectric functions, while the more accurate hybrid nonlocal exchange-correlation functional (HSE) is used to calculate the electronic properties since GGA usually underestimates the band gap for these compounds. As the $^3P_0$-$^3P_2$ spin-orbit splitting is very large for Sn (0.43 eV) and Pb (1.32 eV),[16,17] spin-orbit coupling (SOC) is included in all the calculations.

## III. CRYSTAL STRUCTURES

Experimentally, some earlier studies reported that ABX$_3$ type compounds shows various phases under different temperatures, but in the high temperature phase, all adopt a cubic perovskite structure, as shown in Fig. 1, where a three-dimensional framework of corner-sharing BX$_6$ octahedron is given. [7,8,18-25] For simplicity, we consider only the cubic structure in this work in order to understand the chemical trend of this kind of materials.

We first relax the structures of all the cubic ABX$_3$ type compounds. The obtained lattice constants are shown in Table I. We find that the lattice constants of $ABX_3$ increase with increasing the size of X from Cl, to Br and I. When keeping the B site and X site atoms, the lattice constants of $ABX_3$ will change with the size of A site atom. $CH_3NH_3$ and $NH_2CHNH_2$ have similar size and larger than Cs, so the lattice constant are nearly the same for A = $CH_3NH_3$ and $NH_2CHNH_2$ and it is larger than that with Cs. Comparing the obtained lattice constants with the available experimental results, we can see that our results are in good agreement with experimental data.[7,8,18-22]

## IV. ELECTRONIC AND OPTICAL PROPERTIES

Based on the relaxed structures, we investigate the electronic properties of these compounds, including band structures, density of states and band gaps, etc., as will be discussed below. For the band structures and density of states (DOS), according to our investigation, all the $ABX_3$ compounds possess similar electronic properties, hence we just take $CsSnCl_3$ as an example. Fig. 2 plots the calculated band structure of $CsSnCl_3$. We can see that there exists a direct band gap at the R (0.5, 0.5, 0.5) point. What's more, we find that the electric-dipole transition at R from the valence band maximum (VBM) to the conduction band minimum (CBM) is allowed, which leads to an optical band gap of 1.19 eV. It is worth to mention that, in these compounds, CBM is threefold degenerate if SOC is not included.[9,10] When the SOC effect is considered, CBM splits into a two-fold low-lying $p_{1/2}$ manifold and a four-fold high-lying $p_{3/2}$ manifold (spin degeneracy is counted) with an energy difference of 0.46 eV for $CsSnCl_3$. This splitting can induce a down shift of CBM and decrease band gap by 0.33 eV. For all the cubic $ABX_3$ compounds, the spin-orbit splitting energy of CBM is about 0.46 eV and 1.48 eV for B = Sn and Pb, respectively. As this splitting induces a large decrease of the band gap, the effect of SOC should be included to determine accurately the optical transitions in these systems.

The calculated density of states (DOS) and partial density of states projected on Cs, Sn, Cl atoms are shown in Fig. 3. Contour plots of the total charge density, and charge density at the band edges of $CsSnCl_3$ in the (110) plane are shown in Fig. 4. As we can see, the VBM is mainly the anti-bonding component of the hybridization between Sn *s* states and Cl *p* states, while the CBM is almost a nonbonding state dominated by the Sn *p* orbitals. The electronic levels from Cs are located deep within the valence band and the conduction band, without contribution to the states near the Fermi level. It can be seen clearly from the total charge density (Fig. 4a) that the coupling between Sn and Cl ions is very strong while it's very weak between Cs and Cl ions, which suggests the strong covalent nature of Sn-Cl and strong ionic nature of Cs-Cl (Cs acts just as a charge donor here). In fact, all the $ABX_3$ compounds have similar electronic properties as $CsSnCl_3$, with a direct band gap at R point, the VBM corresponds to the hybridized anti-bonding state of B *s*-X *p* orbitals and the CBM corresponds to the non-bonding B *p* states. B-X is strongly covalent while A-X is strongly ionic and A site atom just plays a role of charge donor.

Since GGA usually underestimates the band gaps, we use the hybrid nonlocal exchange-correlation functional (HSE) to accurately predict the band gaps. The results are given in Table II. The numbers in parentheses are calculated without SOC effect. We can see that, the decrease of the band gap is about 0.35 eV and 1.13 eV for B = Sn and Pb, respectively, when the effect of SOC is included, which suggests that the inclusion of SOC effect is important in determining the band gaps of these systems. As we know, semiconductors with band gap between 1.1 eV and 1.5 eV have the greatest potential for an efficient solar cell.[26-28] From this viewpoint, we can conclude that $CsSnCl_3$, $CsPbBr_3$, $CH_3NH_3SnBr_3$, $CH_3NH_3PbBr_3$, $NH_2CHNH_2SnBr_3$ and $NH_2CHNH_2PbBr_3$ are potential candidates as good solar cell materials since their band gaps are in this range. Besides, we also notice that (i) as the size of A increases, the band gap of $ABX_3$ increases; (ii) as B varies from Sn to Pb, the band gap of $ABX_3$ increases; and (iii) as X changes from Cl to Br to I, the band gap decreases. These chemical trends are important to understand and optimize this type of solar cell absorbers. We explain these trends by the above band structure and DOS analysis as follows: (i) Band gap dependence on the A ion. As an A ion has no contribution to the states near the Fermi surface, it affects the band gaps only indirectly by modifying the lattice constants of $ABX_3$. Since the size of $CH_3NH_3$ and $NH_2CHNH_2$ are similar and larger than Cs, the lattice constants of (CH$_3$NH$_3$)BX$_3$ and (NH$_2$CHNH$_2$)BX$_3$ are very close and larger than CsBX$_3$. Then the larger lattice constant caused by the larger size of A can decrease the strength of the B *s*-X *p* hybridization due to the longer B-X bond. Because the VBM is an anti-bonding state of the hybridization, it is shifted down in energy when the coupling decreases, thus explaining the increased band gap with A varies from Cs to CH$_3$NH$_3$ (NH$_2$CHNH$_2$). (ii) Band gap dependence on the B ion. We consider the atomic-orbital energy levels. Sn 5*s* orbital energy is higher than that of Pb 6*s*, and closer to the X *p* energy levels, so the coupling between Sn *s*-X *p* is stronger than that of Pb *s*-X *p* in the VBM. As VBM is an anti-bonding state of B *s* and X *p*, the stronger coupling between Sn *s* and X *p* upshifts the VBM and decrease the gap of ABX$_3$, which explains the increased band gap with B varies from Sn to Pb. (iii) Band gap dependence on the X ion. Since the VBM contains some X *p* characters and the energy levels of the *p* states of the X ions increases from Cl to I, the VBM upshifts and the band gap becomes smaller from Cl to I. Based on these chemical trends, we can use band structure engineering methods to flexibly tune the electronic properties of this kind of materials, i.e., by forming alloys between them, it is possible to find or design new solar cell absorbers with better performances.

As CsSnCl$_3$, CsPbBr$_3$, CH$_3$NH$_3$SnBr$_3$, CH$_3$NH$_3$PbBr$_3$, NH$_2$CHNH$_2$SnBr$_3$ and NH$_2$CHNH$_2$PbBr$_3$ are more promising materials for solar cells absorber from the viewpoint of band gap, now we focus on these materials and compare their optical properties. In Fig. 5, we plot the imaginary part ($\varepsilon_2(\omega)$) of the calculated dielectric functions for these compounds using PBE method. As the crystals are isotropic (A = CH$_3$NH$_3$ or NH$_2$CHNH$_2$ has little influence to the cubic symmetry of the crystal), all the diagonal components of $\varepsilon_2$ are identical ($\varepsilon_{xx} = \varepsilon_{yy} = \varepsilon_{zz}$), whereas the off-diagonal components are zero. We find that CH$_3$NH$_3$SnBr$_3$ and NH$_2$CHNH$_2$SnBr$_3$ have the best optical properties compared to other compounds as they have the best absorption below 3.5 eV. Besides, they are nontoxic compared to the Pb-based materials. Note that the room-temperature phase of CH$_3$NH$_3$SnBr$_3$ is just the cubic structure,[29,30] which makes CH$_3$NH$_3$SnBr$_3$ more convenient to be used as solar-cell material.

## V. CONCLUSIONS

In conclusion, the electronic and optical properties of cubic $ABX_3$ (A = Cs, $CH_3NH_3$, $NH_2CHNH_2$; B = Sn, Pb; X = Cl, Br, I) type compounds are studied using the first-principles methods which employ the hybrid density functional and include the SOC effect. We find that (i) as the size of A increases, the lattice constant and band gap of $ABX_3$ increases; (ii) as B varies from Sn to Pb, the band gap of $ABX_3$ also increases; (iii) as X ranges from Cl to Br to I, the lattice constant increases and band gap decreases; (iv) all the $ABX_3$ compounds have similar electronic properties, with a direct band gap at R point, the VBM corresponds to the hybridized anti-bonding state of B $s$-X $p$ orbitals and the CBM corresponds to the non-bonding B $p$ states, B-X is strongly covalent while A-X is strongly ionic; (v) $CH_3NH_3SnBr_3$ is a promising nontoxic material for solar cells absorber with a perfect band gap and good optical absorption.

Acknowledgement: The work is partially supported by the Special Funds for Major State Basic Research, National Science Foundation of China (NSFC), Ministry of Education and Shanghai Municipality, FANEDD, Eastern Scholar Program. The calculations were performed in the Supercomputer Center of Fudan University.

*Corresponding Authors: hjxiang@fudan.edu.cn, xggong@fudan.edu.cn

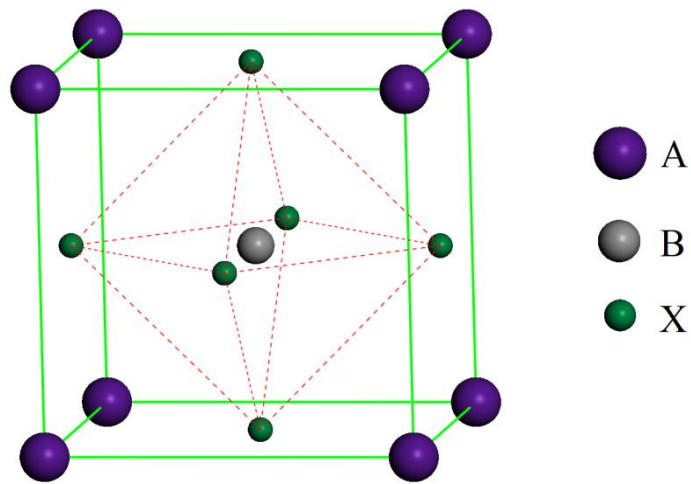

FIG. 1. (color online) Crystal structures of perovskite compounds in the cubic phase. The primitive cell of cubic perovskite $ABX_3$ is consist of corner-linked octahedral of the X atoms, with the B cation at their center and the A cation between them.

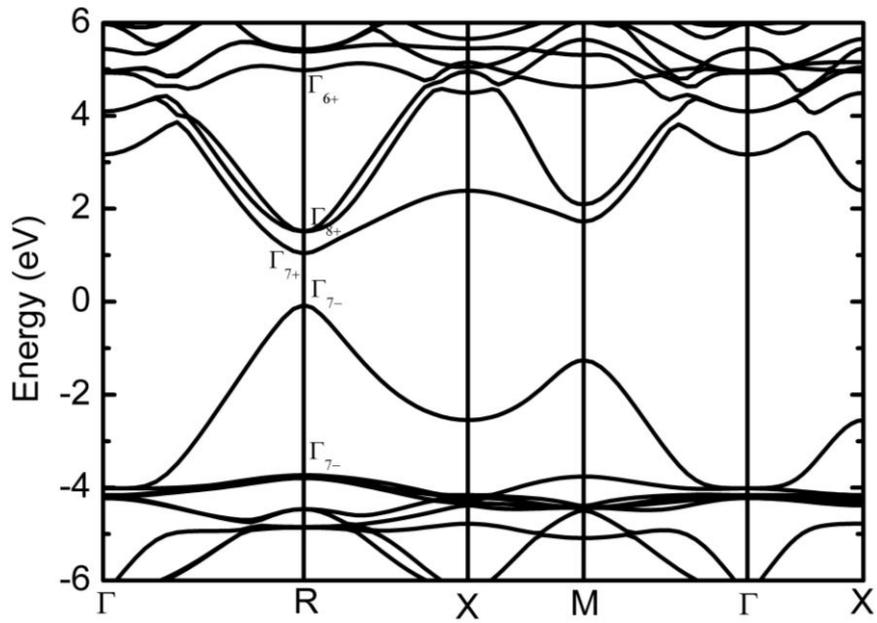

FIG. 2. Calculated band structure for CsSnCl$_3$ using the HSE functional. The energy zero is at the VBM. The symmetries of the bands at the R point are also labeled.

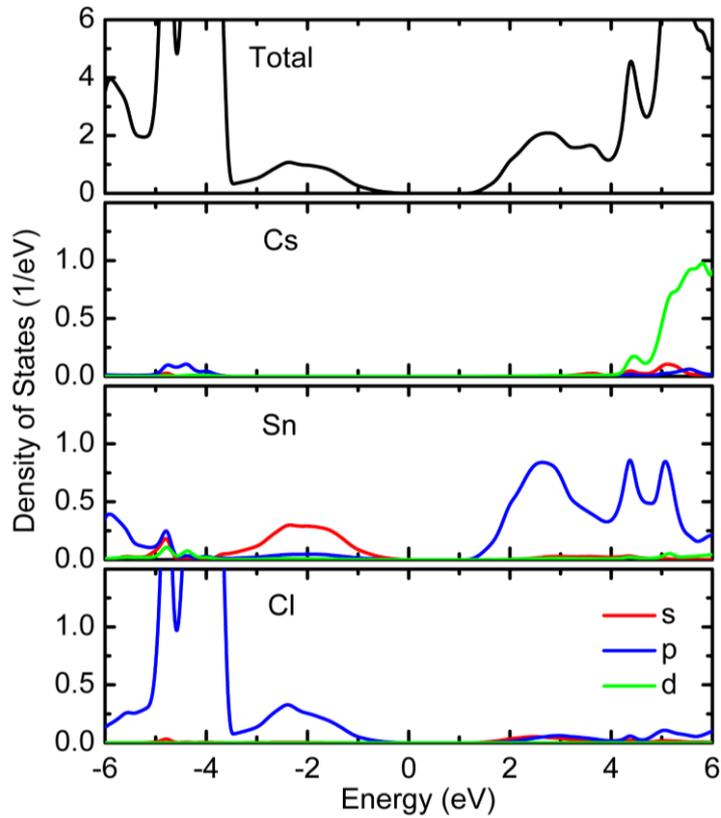

FIG. 3. (color online) Calculated total and partial density of states for CsSnCl$_3$ using the HSE functional. The energy zero is at the VBM.

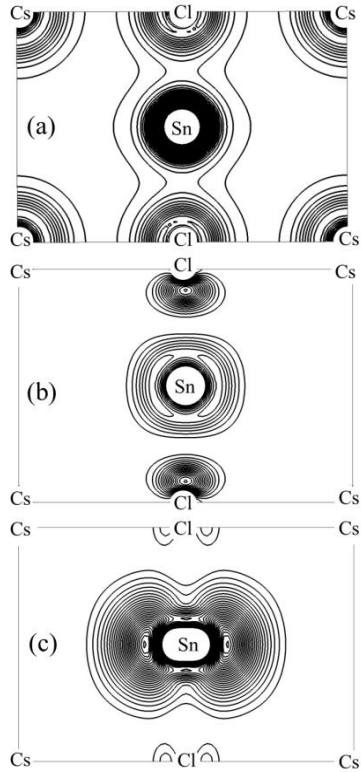

FIG. 4. The contour plots of (a) total charge density, (b) charge density from the VBM and (c) charge density from the CBM for CsSnCl$_3$ in the (110) plane.

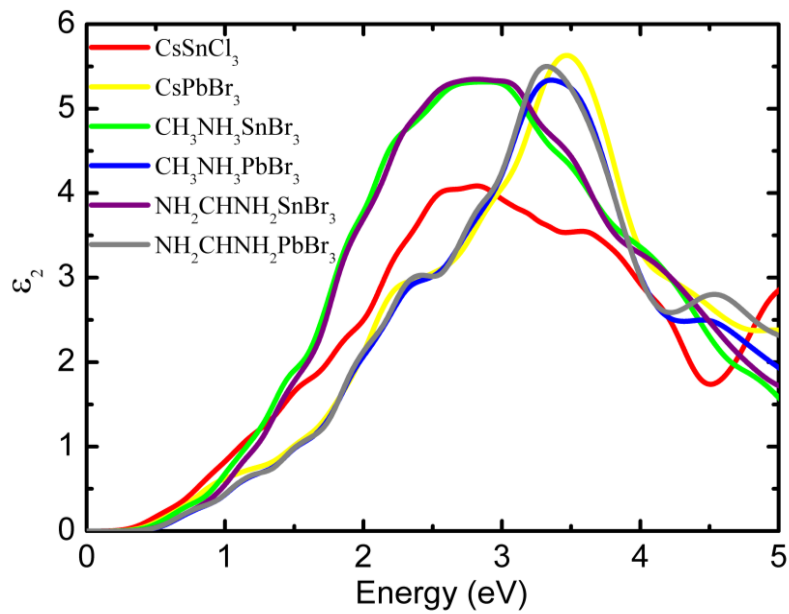

FIG. 5. (color online) Calculated imaginary dielectric functions versus energy by using PBE method for CsSnCl$_3$, CsPbBr$_3$, CH$_3$NH$_3$SnBr$_3$, CH$_3$NH$_3$PbBr$_3$,

NH$_2$CHNH$_2$SnBr$_3$ and NH$_2$CHNH$_2$PbBr$_3$.

| A = Cs | | X | | |
|---|---|---|---|---|
| | | Cl | Br | I |
| B | Sn | 5.61 (5.60[18]) | 5.89 | 6.28 (6.22[19]) |
| | Pb | 5.73 (5.61[21]) | 5.99 | 6.39 |
| A = CH$_3$NH$_3$ | | X | | |
| | | Cl | Br | I |
| B | Sn | 5.90 (5.76[18]) | 6.10 (5.89[8]) | 6.41 (6.24[7]) |
| | Pb | 5.80 (5.68[20]) | 6.10 (5.90[20]) | 6.46 (6.33[20]) |
| A = NH$_2$CHNH$_2$ | | X | | |
| | | Cl | Br | I |
| B | Sn | 5.92 | 6.13 | 6.46 (6.32[22]) |
| | Pb | 5.81 | 6.09 | 6.47 |

TABLE I. Calculated lattice constants (in Å) of ABX$_3$ compounds (with A = Cs, CH$_3$NH$_3$, NH$_2$CHNH$_2$; B = Sn, Pb; X = Cl, Br, I). The available experimental results are shown in parentheses.

| A = Cs | | X | | |
|---|---|---|---|---|
| | | Cl | Br | I |
| B | Sn | 1.19 (1.52) | 0.80 (1.14) | 0.49 (0.87) |
| | Pb | 1.83 (2.92) | 1.32 (2.41) | 0.86 (2.00) |
| A = CH$_3$NH$_3$ | | X | | |
| | | Cl | Br | I |
| B | Sn | 1.94 (2.27) | 1.31 (1.66) | 0.75 (1.13) |
| | Pb | 1.98 (3.08) | 1.48 (2.61) | 0.95 (2.10) |
| A = NH$_2$CHNH$_2$ | | X | | |
| | | Cl | Br | I |
| B | Sn | 1.98 (2.31) | 1.37 (1.72) | 0.84 (1.23) |
| | Pb | 2.00 (3.10) | 1.46 (2.59) | 0.96 (2.11) |

TABLE II. Calculated band gaps (in eV) of ABX$_3$ compounds (with A = Cs, CH$_3$NH$_3$, NH$_2$CHNH$_2$; B = Sn, Pb; X = Cl, Br, I) using the HSE functional. The results in parentheses are obtained without SOC effect.